\documentclass[aps,prl,twocolumn,amsmath,amssymb,showpacs,superscriptaddress]{revtex4-1}
\usepackage{graphicx}% Include figure files
\usepackage{subcaption} 
\usepackage{dcolumn}% Align table columns on decimal point
\usepackage[mathlines]{lineno}
\usepackage{physics}
\usepackage{hyperref}
\usepackage{multirow}
\usepackage{bm}% bold math
\usepackage{epstopdf}
\usepackage{epsfig}
\usepackage{bbold}
\usepackage[normalem]{ulem}
\usepackage{caption}
\usepackage[dvipsnames]{xcolor}
\usepackage{comment}
\usepackage{amsmath}
\newcommand{\note}[1]{\textcolor{red}{ #1}}

\begin{document}
%%%%%%%%%%%%%%%%%%%title%%%%%%%%%%%%%%%%%%%
\title{Particle-like topologies of light in turbulent complex media}

\author{Danilo Gomes Pires}
\thanks{These authors contributed equally to this work.}
\affiliation{Department of Electrical and Computer Engineering, Duke University, Durham, USA}

\author{Vasilios Cocotos}
\thanks{These authors contributed equally to this work.}
\affiliation{School of Physics, University of the Witwatersrand, Private Bag 3, Wits 2050, South Africa}

\author{Cade Peters}
\affiliation{School of Physics, University of the Witwatersrand, Private Bag 3, Wits 2050, South Africa}

\author{Natalia M. Litchinitser}
\email[email:]{natalia.litchinitser@duke.edu}
\affiliation{Department of Electrical and Computer Engineering, Duke University, Durham, USA}

\author{Andrew Forbes}
\email[email:]{andrew.forbes@wits.ac.za}
\affiliation{School of Physics, University of the Witwatersrand, Private Bag 3, Wits 2050, South Africa}

\date{\today}

%%%%%%%%%%%%%%%%%%%%%abstract%%%%%%%%%%%%%%%%%%%%
\begin{abstract}
\noindent The basic building blocks of many forms of optical topologies are particle-like singularities in phase and polarisation, giving rise to lines of darkness that weave complex threads in 3D space. Although known for half a century since seminal work on dislocations in wave trains, their behaviour in complex media remains under debate, especially with respect to their relative stability. Here we show that polarisation and phase vortices behave identically in one-sided turbulent complex channels. We perform complementary numerical and experimental studies using atmospheric turbulence as a test case, demonstrating agreement and equivalent dynamics. Our work addresses open questions on optical topologies and will be relevant to their harnessing for applications such as sensing, communication, imaging, and information transfer in noisy or complex environments. 
\end{abstract}

\maketitle

Since the identification of wavefront dislocations by J. Nye and M. Berry \cite{nye1974dislocations}, optical singularities have served as compact markers of field structure, capturing where a chosen field descriptor loses a unique definition \cite{soskin2001singular}. In the scalar setting, an example is the optical vortex. It is defined as a point (or line) of vanishing complex amplitude around which the wavefront wraps, providing a natural route to beams carrying orbital angular momentum (OAM) \cite{franke202230}. Within this broader class of OAM states, the Laguerre-Gaussian (LG) modes occupy a particular position of interest. They provide a widely used modal basis whose azimuthal order sets a helical wavefront and radial order controls the ringed intensity profile, making them a standard platform for generating, combining, and diagnosing OAM content \cite{fontaine2019laguerre,schulze2012wavefront,rafayelyan2017laguerre}. By arranging singularities through controlled superposition of such modes, one can assemble extended topological architectures, ranging from vortex lattices \cite{zhu2021optical,dreischuh2002generation} to knotted nodal filaments \cite{dennis2010isolated,leach2005vortex,pires2025stability}. In vectorial fields, analogous design principles enable polarisation textures whose topology is encoded not only in the spatial distribution but also in how the local internal state is disposed across the beam cross-section \cite{larocque2018reconstructing,shen2024optical,guo2026topological,lei2025topological,pires2025structuring}.
%spanning Bessel beams \cite{durnin1987diffraction,mcgloin2005bessel,khonina2020bessel}, perfect-vortex constructions \cite{ostrovsky2013generation,fu2016perfect}, and spatiotemporal vortices \cite{bliokh2012spatiotemporal,jhajj2016spatiotemporal,hancock2019free,chong2020generation}, 

A crucial separation must be drawn between phase singularities, known as defects of a complex scalar field that anchor OAM-carrying beams, and polarisation singularities, arising when the local polarisation state becomes indeterminate even though the total intensity may remain finite. Because these singularities reside in different field descriptors, the literature contains competing expectations of their resilience under external perturbations. One viewpoint emphasizes the quantized nature of the phase winding in OAM beams, arguing that the associated defect can shift or split but cannot be continuously eliminated, suggesting enhanced robustness to weak distortions \cite{meglinski2024phase,bouchal2002resistance,meglinski2024phase}. An opposing perspective notes that polarisation textures can persist when the channel primarily scrambles spatial modes while leaving the polarisation sector effectively coherent, so that vectorial structure can remain identifiable even as the scalar wavefront becomes highly corrugated \cite{lochab2018designer,yu2025640,lochab2019propagation}. Therefore, reported conclusions depend sensitively on what quantity is used to define \textit{stability}, which can be through the defect location \cite{cui2014influence,gu2013statistics}, OAM spectrum \cite{lavery2018vortex,li2016compensation,ren2014adaptive}, or global topological descriptors \cite{wang2024topological,peters2023invariance}. This ambiguity is particularly pronounced in atmospheric turbulence and other random environments, where the same disturbance can simultaneously broaden the OAM spectrum, induce vortex migration, and deform polarisation features, leading to a nontrivial comparison between phase- and polarisation-defined singular structures, often contradictory in some cases.

Here we address these contradictory perspectives by revealing the dynamics of both phase and polarisation singularities in one-sided complex channels, using atmospheric turbulence as an example. We outline theoretically how a one-sided channel acts on singularities in a manner that can be generalised to any degree of freedom.  We confirm this experimentally by subjecting both phase and polarisation singularities to the same atmospheric turbulence conditions and quantify how the associated dislocation behaves in the far-field by inferring their wandering statistics and their sensitivity to increasing perturbation strength. This approach establishes an experimentally accessible benchmark for resolving prior ambiguities in comparative robustness and provides a concrete basis for designing structured-light encodings whose stability is set by the intended field descriptor.

%\begin{figure*}[t!]
 %   \centering
  %  \includegraphics[width=\textwidth]{Figures/Figure 1.png}
   % \caption{%Experimental setup to generate and measure (a) vector and (b) scalar beams carrying polarisation and phase singularities, respectively. In both arrangements, the component labels mean spatial light modulator (SLM), lenses (L), iris apertures (Ap), mirrors (M), half-wave plates (HWP), quarter-wave plate (QWP), and charge-coupled device (CCD), which is a polarisation-sensitive camera. The insets show examples of the holograms used for both experimental arrangements.}
 %   }
  %  \label{fig:concept}
%\end{figure*}

\begin{figure*}[t!]
    \centering
    \includegraphics[width=\textwidth]{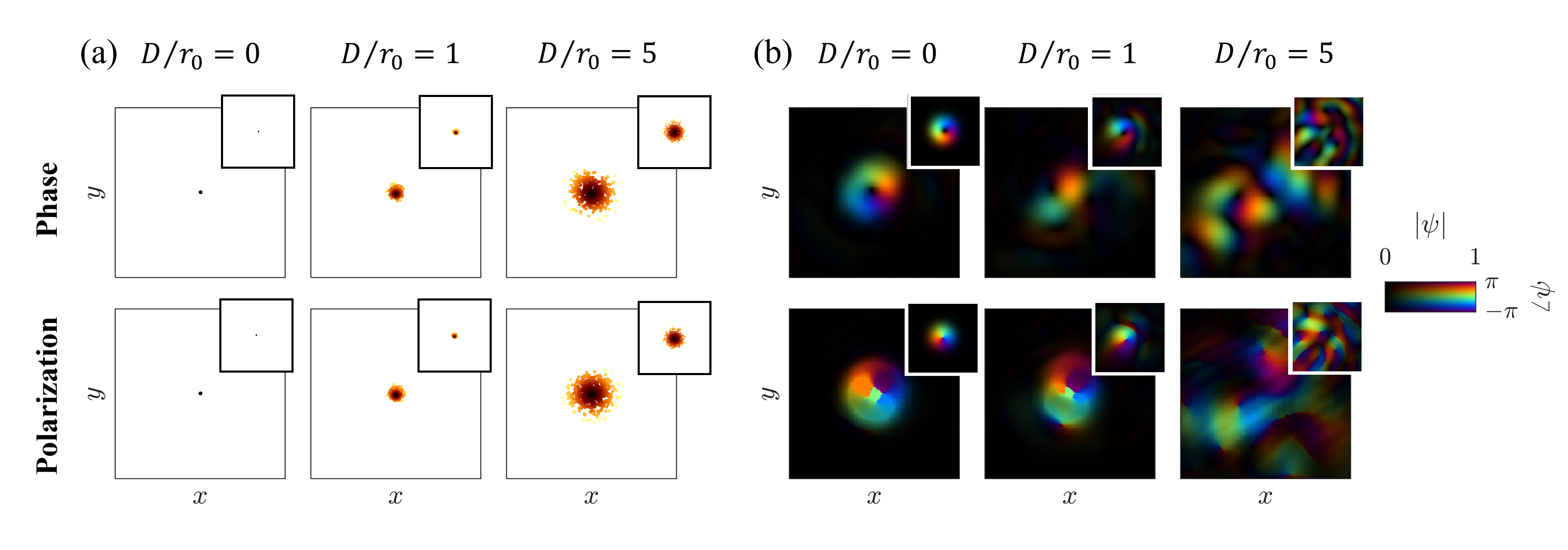}
    \caption{(a) Singularity constellation distributions and (b) typical complex fields obtained for turbulent strengths $D/r_0=0,1,5$, respectively distributed across the rows. Within each panel, the top (bottom) row refers to the $\ell=1$ vortex encoded in the phase (polarisation) distribution. Each inset refers to the simulation distributions, while the rest corresponds to their respective experimental result. The marker colours in panel (a) represent their respective distance from the constellation centroid. In panel (b), brightness represents the total amplitude $|\psi|$, while the colours refer to the phase $\angle \psi$ for the scalar vortex and to the Stokes field phase $\text{atan}(S_2/S_3)$ for the polarisation vortex.}
    \label{fig:sing}
\end{figure*}

To implement the phase-singular beam, we take a conventional scalar vortex beam given by a single Laguerre Gaussian (LG) mode with azimuthal index $\ell$ and zero radial order in a fixed polarisation state $\hat{\textbf{e}}$, given by $\psi_{\text{ph}}(\textbf{r}_\perp)=u_\text{LG}^\ell(\textbf{r}_\perp) \hat{\textbf{e}}$, where the separability ensures that we have only a single spatial degree of freedom that can be acted on. For the polarisation-encoded dislocation, we define a two-component transverse field in the linear polarisation basis as a coherent superposition of a fundamental Gaussian mode ($\ell=0$) in the horizontal channel $\hat{\textbf{e}}_\text{H}$ and a single LG vortex mode in the vertical channel $\hat{\textbf{e}}_\text{V}$, both evaluated at the same longitudinal plane $z=0$ across the transverse coordinates $\textbf{r}_\perp$. In our notation, the polarisation-singular field is written compactly as
\begin{equation}
    \psi_{\text{pol}}(\textbf{r}_\perp)=u_\text{LG}^{\ell = 0}(\textbf{r}_\perp) \hat{\textbf{e}}_\text{H}+u_\text{LG}^\ell(\textbf{r}_\perp) \hat{\textbf{e}}_\text{V},
\end{equation}
where $u_\text{LG}^\ell$ stands for the same LG mode previously defined. This field has two non-separable degrees of freedom (DoF) and ensures that the targeted singular behaviour is carried by the spatially varying inter-component phase relation rather than by the phase of a single scalar field, while keeping the modal content minimal and unambiguous. Now, both cases share the same OAM order but differ in whether the dislocation is encoded in a one-component phase structure or in a two-component polarisation superposition (polarisation singularity). The experimental setups used to generate both scalar and vector beams and simulate the effects of atmospheric turbulence are shown in the Supplementary Materials. The setups were carefully designed to ensure both singularity types were generated in a comparable manner and experienced identical realizations of the perturbing channel.
% Should we briefly emphasize that the setups were made as similar as possible for the fair comparison?

Suppose we wish to pass these two beams through a one-sided complex channel acting only on degree of freedom $B$ while leaving the other ($A$) unaltered. In such a case the action of the channel is given by $\mathbb{1}_A \otimes T_{B}$, where $\mathbb{1}_B$ is the identity matrix and $T_B$ is the channel operator. In what follows we will use atmospheric turbulence as our example so that the subscripts $A,\, B$ refer to the polarisation and spatial DoFs respectively, but stress that these can be switched or altered to other DoFs for other complex channels. We select atmospheric turbulence as an exemplary one-sided channel (since it is non-birefringent) because of its fundamental and practical importance. Such free-space channels often deteriorate the original transverse structure of light, as random refractive-index inhomogeneities imprint spatially varying optical-path delays, redistribute energy across the beam profile, and thereby perturb the dislocation geometry that carries the information or the topological descriptor. Here, the key question is not whether a singular pattern can be identified in an idealized field, but whether it remains traceable and statistically predictable after propagation through a stochastic medium, particularly when the singular structure is embedded either in a single-component wavefront or in a two-component polarisation superposition. To model atmospheric turbulence, we adopt the standard Kolmogorov description, which treats refractive-index fluctuations as a statistically homogeneous and isotropic random process with a scale-invariant cascade over an inertial range bounded by inner and outer lengths \cite{peters2025structured}. This framework relates turbulence strength and correlation scales to the induced wavefront distortions and intensity modulation, and forms the basis for the numerical phase-screen simulations used here. Details on the Kolmogorov model, turbulent phase screen generation, and experimental calibration are presented in the Supplementary Materials.

A direct way to compare the dynamical responses of the phase and polarisation vortices in turbulence is to quantify the wandering of their singularities, namely, the stochastic displacement of the deflect coordinates across an ensemble of distorted realizations. Because the singularity position is a local observable that can be defined for both scalar and polarisation vortices, its statistics provide a common metric for comparing the sensitivity of the corresponding topological descriptors to the same random medium. In practice, for each realization and both cases, we first determine the beam's centre of mass from the perturbed intensity distribution, which defines a robust reference point even when the pattern is distorted. For the $\ell=1$ case, a single singularity is located close to the beam centre. For higher orders ($\ell=2$ and $\ell=3$), we identify  $n=\ell$ singularities clustered around the centre-of-mass region, reflecting the expected local multiplicity near the beam core under perturbation. Aggregating these coordinates across all realizations yields a spatial occupancy map of singularity locations for each $\ell$ and each turbulence strength, parametrized by $D/r_0$. Here, $D=2w_0$ stands for the beam diameter with respect to its waist size, while $r_0$ refers to the Fried parameter. Throughout this work, a beam waist of $w_0=1$ mm was considered for both polarisation and phase vortices.

\begin{figure}
    \centering
    \includegraphics[width=\linewidth]{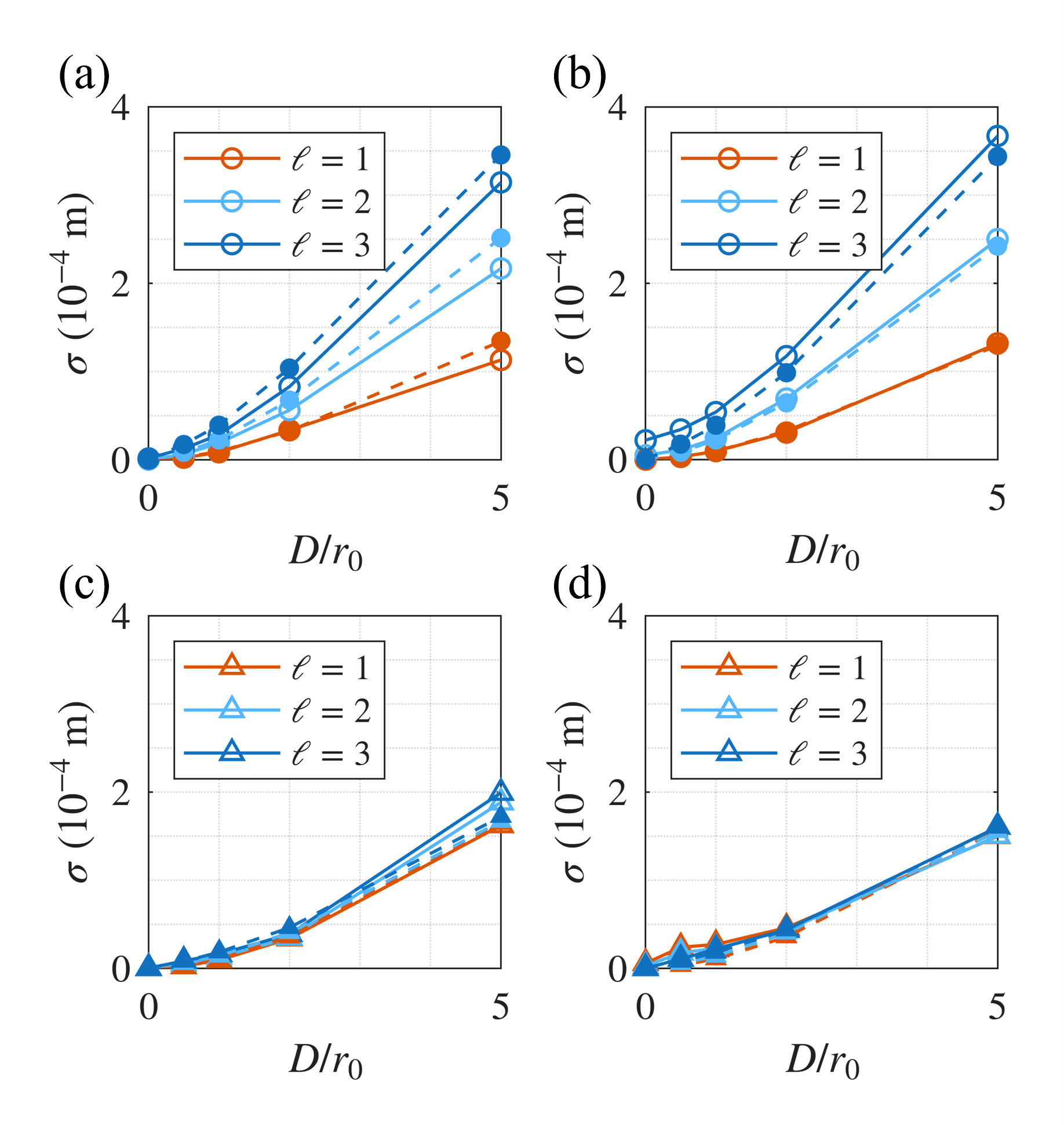}
    \caption{Positional deviation radius $\sigma$ as a function of the turbulence strength $D/r_0$ for (a) the phase singularity case and (b) the polarisation singularity case. The same experiment was repeated with corrected beam waists (see main text), (c) for the phase singularity case and (d) for the polarisation singularity case. All plots show the outcomes for topological charges $\ell=1,2,3$, where the solid lines and hollow markers refer to the simulation results, while dashed lines and filled markers represent experiments.}
    \label{fig:stddev}
\end{figure}

Figure~\ref{fig:sing} (a) shows a comparison between the typical constellation of singularity locations for $\ell=1$ for the cases of both phase and polarisation vortices and turbulence strengths $D/r_0=0,1,$ and $5$, respectively. Similarly, Fig.~\ref{fig:sing} (b) presents examples of the complex fields of scalar and vector beams obtained from simulations and experiments for $\ell=1$ with the same turbulence strengths as in Fig.~\ref{fig:sing} (a). An ensemble of 500 realizations for each turbulence strength was used to obtain each constellation map, and details on the numerical simulations are provided in the Supplementary Materials. To further verify the similarities between the two cases observed in Fig.~\ref{fig:sing} (a), systematically compare wandering statistics across $\ell=1,2,3$ for various turbulence levels by extracting their corresponding positional deviation radius $\sigma=\sqrt{\sigma^2_x+\sigma_y^2}$, where $\sigma^2_{x,y}$ refers to the variance along the $x,y$ directions. Across the full range of turbulence strengths explored, $\sigma$ exhibits the same qualitative evolution for both types of vortices, as shown in Fig.~\ref{fig:stddev} (a) and (b). Here, the distributions for phase and polarisation singularities, respectively, broaden monotonically with increasing $D/r_0$, and, at a fixed turbulence level, the spread increases systematically as the topological charge changes from $\ell=1$ to $\ell=2$ and $\ell=3$. This agreement shows that vortex wandering is set by turbulence-induced distortions, independent of whether the defect appears as a phase or polarisation vortex. Once the beam is referenced to its fluctuating centre of mass, the residual motion of the nearest singularity (or, for higher charge, the cluster of $n=\ell$ singularities) follows the same turbulence-driven statistics in both cases, leading to a comparable deviation scaling within the explored parameter space.

Generally, in a random but non-birefringent atmosphere, turbulence acts primarily as a polarisation-independent complex perturbation to the transverse field \cite{ndagano2017creation,cvijetic2010polarization}. Consequently, the polarisation component carrying the LG structure (in our case, vertical polarisation state) experiences essentially the same wavefront deterioration and transverse wavevector fluctuations whether it is analysed as a scalar field or embedded within a two-component polarisation superposition. The polarisation singularity is then bound to the evolution of that LG component, since the Gaussian mode counterpart serves mainly as a smooth reference, not introducing a competing singular structure. As a result, the singularity position extracted from the polarisation descriptor inherits the same random displacements as the underlying scalar LG vortex. 
%This is established analytically in the Supplementary Materials using two independent approaches. In a one-sided channel description, turbulence is modelled as a polarisation neutral transformation $\mathbb{1}_A \otimes T_{B}$, where $\mathbb{1}_B$ acting on the polarisation $(A)$ and spatial $(B)$ domains, which forces the phase vortex and the polarisation singularity tied to the same LG component to undergo identical transverse displacements. Alternatively, a wave optics informed treatment based on the Huygens-Fresnel propagation integral of a phase-distorted vectorial beam confirms the same behaviour, showing identical spatial wandering of the two singularities. These theoretical descriptions are general, and are valid for various systems ranging from non-unitary transformations to chiral media \cite{nape2022revealing}. 
This is established analytically in the Supplementary Materials. In a one-sided channel description, turbulence is modelled as a polarisation neutral transformation $\mathbb{1}_A \otimes T_{B}$, where $\mathbb{1}_B$ acting on the polarisation $(A)$ and spatial $(B)$ domains, which forces the phase vortex and the polarisation singularity tied to the same LG component to undergo identical transverse displacements. This can be shown explicitly for atmospheric turbulence using a wave-optics treatment, based on the Huygens-Fresnel propagation integral of a phase-distorted vectorial beam, which exhibits identical spatial wandering of the two singularity types. The theoretical treatment is general, and is valid and can be adapted to a variety of systems ranging from non-unitary transformations to chiral media \cite{nape2022revealing}.

The systematic growth of total deviation radius $\sigma$ with $|\ell|$ is likewise consistent with the greater susceptibility of higher-order vortices to random mixing into lower-order components. This is due to the instability of a multi-charge core, typically $|\ell|>1$, which breaks into multiple unit-charge singularities whose relative positions are determined by the external perturbation \cite{dennis2009singular,shen2019optical}. Both effects expand the ensemble of accessible singularity positions, producing broader constellation maps and consequently larger positional deviations for $|\ell|>1$ in both scalar and polarisation vortex cases.

When the deviation analysis is repeated using corrected-scaling LG modes \cite{cocotos2025orbital}, implemented by adjusting the embedded beam waist as $w_0\rightarrow w_0/\sqrt{|\ell|+1}$, the spatial deviation radius for $\ell=1,2,3$ increases at nearly the same rate as the turbulence strength $D/r_0$ for both scalar and polarisation vortices. Figure~\ref{fig:stddev} (c) and (d) shows the deviation curves as a function of the turbulence strength. This behaviour contrasts with the uncorrected case, where the apparent $\ell$-dependence is amplified because higher-charge LG beams are typically larger and, therefore, sample a larger effective turbulence aperture. This result is consistent with prior observations that, once the beam size is decoupled from the topological charge, turbulence-induced OAM crosstalk becomes largely independent of the initial $\ell$, as the relevant control parameter is the turbulence strength referenced to a common spatial scale rather than the charge itself. \cite{klug2021orbital}. In our case, the same charge-independent broadening emerges in the positional statistics of the tracked singularities. As the LG-bearing component experiences the same polarisation-insensitive random phase perturbation in a non-birefringent channel, normalizing the mode size forces all $\ell$ modes to experience comparable distortions, and the resulting wandering deviation radius becomes approximately charge-invariant within experiments and simulation for both types of vortices.

\begin{figure}[t!]
    \centering
    \includegraphics[width=0.6\linewidth]{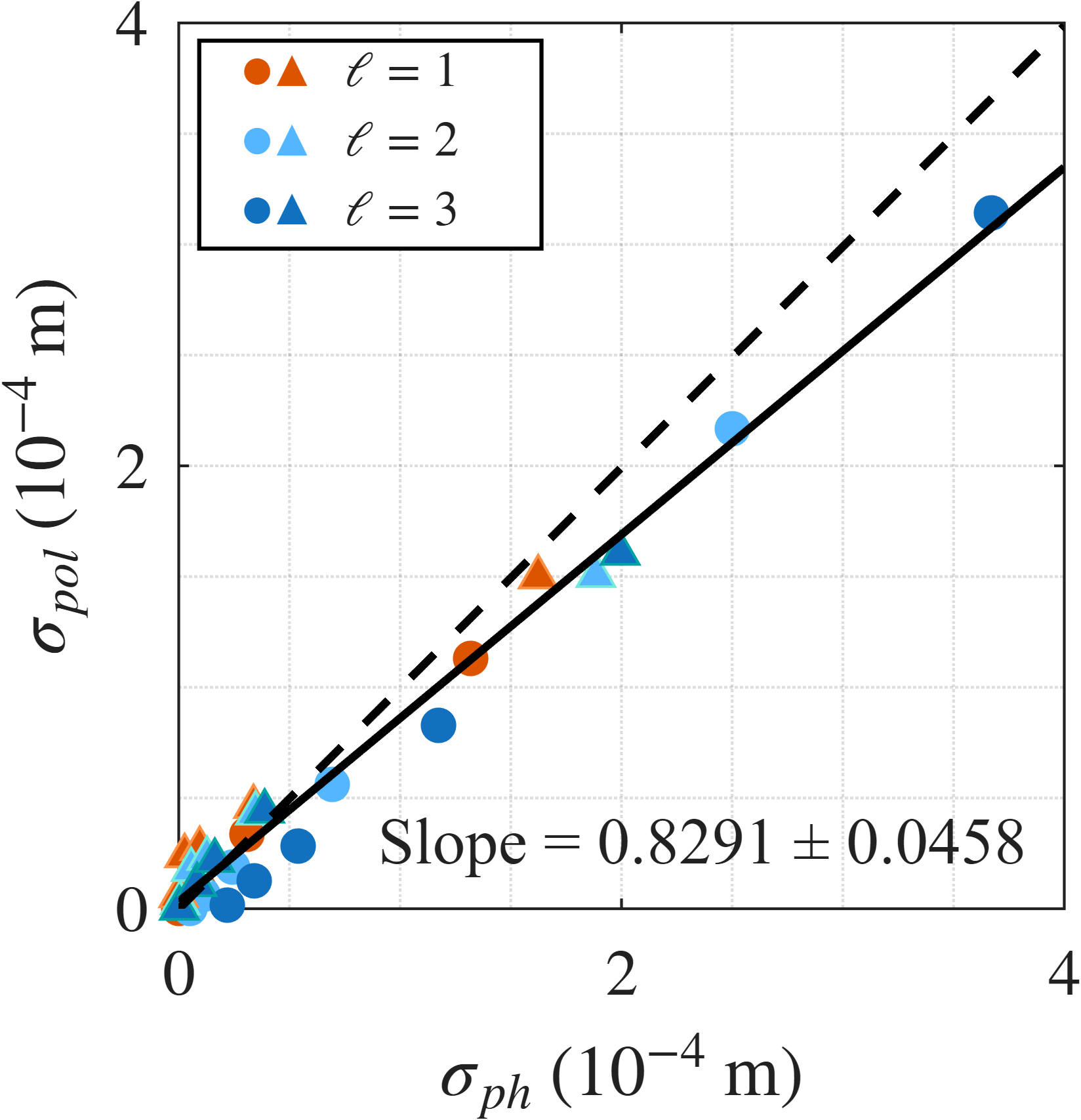}
    \caption{Experimentally retrieved positional deviation radius for the polarisation ($\sigma_{pol}$) and phase ($\sigma_{ph}$) vortices generated using the LG mode with uncorrected (circles) and corrected (triangles) beam waist with topological charges $\ell=1,2,3$. The black solid line refers to the experimental linear fit, with slope $0.8291 \pm 0.0458$. The dashed line corresponds to the theoretical linear fit, with slope $0.9971 \pm 0.0190$.}
    \label{fig:final}
\end{figure}

To further condense the comparison into a single, vortex type-agnostic diagnostic, we directly plot the experimentally measured positional deviations $\sigma$ for the polarisation-encoded singularities against the corresponding total deviation radius for the phase-encoded singularities, treating each turbulence condition and $\ell$ state as a point in a two-parameter space. In this representation, perfect equivalence corresponds to the identity line, and the fitted slope quantifies the relative response of the two vortex types to the same perturbations. Across the explored range of $D/r_0$, the results follow a linear trend with slope near unity, indicating that the ensemble broadening of singularity positions is statistically equivalent in the two domains. Figure~\ref{fig:final} summarizes these findings for both LG modes with uncorrected and corrected beam waists in experiments, with the inferred slope of $0.8291 \pm 0.0458$ in close agreement with theory which predicts a slope of $0.9971 \pm 0.0190$. Deviations from theory primarily arise from finite ensemble size and experimental noise in singularity localization. Taken together, these results provide direct evidence of equivalent sensitivity to turbulence for phase and polarisation  vortices. 

In summary, we performed controlled experimental and numerical studies that place phase and polarisation singularities on equal footing through matched field generation, sampling, and intensity-referenced localization. Under identical turbulence conditions, the measured and simulated singularity constellation maps yield closely aligned positional deviation radii for scalar and polarisation vortices, demonstrating that the dominant wandering statistics are set by the turbulence acting on the shared structured component rather than by the vortex type. These results resolve longstanding ambiguity by identifying the conditions under which phase and polarisation singularities exhibit equivalent stability metrics.  Our work provides a quantitative basis for selecting and normalizing structured light encodings for free-space optical links, turbulence-aware metrology, and environmental sensing where reliable singularity tracking, rather than ideal field reconstruction, is the operational figure of merit \cite{chen2025weather,cheng2025metrology}.

\clearpage
\appendix

\setcounter{section}{0}
\setcounter{figure}{0}
\setcounter{table}{0}
\setcounter{equation}{0}
\setcounter{footnote}{0}
\renewcommand{\thesection}{S\arabic{section}}
\renewcommand{\thefigure}{S\arabic{figure}}
\renewcommand{\thetable}{S\arabic{table}}
\renewcommand{\theequation}{S\arabic{equation}}

\section*{Supplementary: The position correspondence of phase and polarisation singularities in single-sided channels} 
Suppose we wish to pass these two beams through a one-sided complex channel acting only on degree of freedom $B$ while leaving the other ($A$) unaltered. In such a case the action of the channel is given by $\mathbb{1}_A \otimes T_{B}$, where $\mathbb{1}_B$ is the identity matrix and $T_B$ is the channel operator. In what follows we will use atmospheric turbulence as our example so that the subscripts $A,\, B$ refer to the polarisation and spatial degrees of freedom respectively, but stress that these can be switched or altered for other complex channels \cite{nape2022revealing}.

Let us generate a vectorial beam of the form
\begin{equation}
    \ket{\Psi_{\text{in}}}=\ket{e_1}_A\ket{u_1}_B + \ket{e_2}_A\ket{u_2}_B,
\end{equation}
where $A$ and $B$ denotes the polarisation and transverse spatial modes. A one-sided channel acting only on $B$ produces
\begin{equation}
\begin{split}
        \ket{\Psi_{\text{out}}}=&(\mathbb{1}_A \otimes T_{B})\ket{\Psi_{\text{in}}}\\
        =&\ket{e_1}_A\ket{v_1}_B + \ket{e_2}_A\ket{v_2}_B,
        \label{psiout}
\end{split}
\end{equation}
with $\ket{v_j}=T_B\ket{u_j}$.

In general, phase singularities in scalar fields are its wave dislocations, located where the complex field vanishes as $u(\textbf{r}_s)=0 \Rightarrow \Re(u)=\Im(u)=0$, with a phase circulation around $\mathbf{r}_s$. On the other hand, polarisation singularities can be found from the Stokes field $Q(\textbf{r})=u_1(\textbf{r})u_2^*(\textbf{r})$, which carries the relative phase between the polarisation components. For this case, its zeros satisfy $Q(\textbf{r})=0$, meaning that $\Re(u_1)=\Im(u_1)=0$ or $\Re(u_2)=\Im(u_2)=0$. This means that a location where the polarisation ellipse orientation becomes undefined is fixed by the non-vanishing of one complex component, and it is co-located with a phase singularity of that component.

Now, since both polarisation components experience the same operator $(\mathbb{1}_A \otimes T_{B})$, the Stokes field can be expressed at the detector plane $\textbf{q}$ as
\begin{equation}
    Q^{\text{out}}(\textbf{q})=(T_Bu_1)(\textbf{q}) (T_Bu_2^*)(\textbf{q}),
\end{equation}
with singularities satisfying $\Re(T_Bu_1)=\Im(T_Bu_1)=0$ or $\Re(T_Bu_2)=\Im(T_Bu_2)=0$. This means that a given polarisation singularity is extracted from the polarisation-resolved complex amplitudes. The C-points occur at locations where one polarisation component vanishes while the other remain finite, and more generally polarisation singular features are induced by the singular behaviour in the underlying scalar components.

For instance, let us select a set of C-points $\textbf{q}_c$ defined by zeros of the $\ket{e_1}$ component in Eq. \ref{psiout} at the detector plane $\textbf{q}$ as
\begin{equation}
    \textbf{q}_c \in \{\textbf{q}:\Re(T_Bu_1)(\textbf{q})=\Im(T_Bu_1)(\textbf{q})=0 \}.
    \label{setqc}
\end{equation}
On the other hand, we get for the scalar field
\begin{equation}
    \textbf{q}_v \in \{\textbf{q}:\Re(T_Bu)(\textbf{q})=\Im(T_Bu)(\textbf{q})=0 \}.
    \label{setqv}
\end{equation}
In our scenario, the scalar field is written as $u_\text{ph}(\textbf{r})=u_\text{LG}^\ell(\textbf{r})\ket{e_1}$, and expressed at the detector plane as 
\begin{equation}
    u_\text{ph}^\text{out}(\textbf{q})=(T_Bu_\text{LG}^\ell)(\textbf{q})\ket{e_1}.
\end{equation}
Here, the Laguerre-Gaussian (LG) modes are defined by
\begin{eqnarray}
   u_\text{LG}^{\ell }(r,\phi) &=& \sqrt{\frac{2p!}{w^2_0\pi(p+|l|)!}}
          \left(\frac{r\sqrt{2}}{w_0}\right)^{|\ell|}
          L_p^{|\ell|}\left(\frac{2r^2}{w^2_0}\right) \nonumber \\
          &\times& e^{-r^2/w^2_0}e^{-i\ell\phi}\,,
          \label{eq:LGdef}
\end{eqnarray} 
where $\textbf{r}=(r,\phi)$ is the transverse spatial coordinate, $w_0$ is the second moment radius of the embedded fundamental Gaussian envelope, $p$ is the radial index and $\ell$ is the azimuthal index that endows the beam with an orbital angular momentum (OAM) of $\ell\hbar$ per photon. For convenience, we set $p=0$ for all simulated and experimental measurements. The second moment beam radius for LG modes is given by $w = w_0 \sqrt{2p + |\ell| + 1}$, and if $\ell=p=0$, Eq. \ref{eq:LGdef} reduces to the fundamental Gaussian beam.

Similarly, the vector field carrying a polarisation singularity is expressed as $u_\text{pol}(\textbf{r})=u_\text{LG}^\ell(\textbf{r})\ket{e_1}+u_\text{LG}^0(\textbf{r})\ket{e_2}$, and at the detector plane as
\begin{equation}
    u_\text{pol}^\text{out}=(T_Bu_\text{LG}^\ell)(\textbf{q}))\ket{e_1}+(T_Bu_\text{LG}^0)(\textbf{q})\ket{e_2}.
\end{equation}
This representation agrees with Eq. \ref{setqc} as the Gaussian component does not possess any singularities, $\Re(u_\text{LG}^0)=\Im(u_\text{LG}^0) \neq 0$. Since the single-sided channel does not introduce any additional singularities, the relation $\Re(T_Bu_\text{LG}^0)=\Im(T_Bu_\text{LG}^0)=0$ never holds. Thus, the set of singular points in Eq. \ref{setqc} and \ref{setqv} becomes, respectively,
\begin{equation}
    \textbf{q}_c \in \{\textbf{q}:\Re(T_Bu_\text{LG}^\ell)(\textbf{q})=\Im(T_Bu_\text{LG}^\ell)(\textbf{q})=0 \},
\end{equation}
and
\begin{equation}
    \textbf{q}_v \in \{\textbf{q}:\Re(T_Bu_\text{LG}^\ell)(\textbf{q})=\Im(T_Bu_\text{LG}^\ell)(\textbf{q})=0 \}.
\end{equation}
Alternatively, this means that $\textbf{q}_v=\textbf{q}_v$. Within a one-sided complex channel, the C-points locations defined by the $\ket{e_1}$ polarisation component and the phase vortex locations of the scalar beam are identical. Consequently, any wandering observable built purely from singularity positions, such as the quantities studied in this framework, must coincide.

\noindent Alternatively we may invoke a wave optics approach.  Consider an arbitrary vector beam,
\begin{equation}
    \mathbf{U}(\textbf{r})=g(\textbf{r}) \hat{\textbf{e}}_\text{R}+u(\textbf{r}_\perp) \hat{\textbf{e}}_\text{L},
\end{equation}
where $\hat{\textbf{e}}_{\text{R}(\text{L})}$ is a unit vector indicating right (left) circular polarisation, $g(\textbf{r})$ is the fundamental Gaussian beam and $u(\textbf{r})$ is a structured field containing a finite number of spatially distributed phase singularities. A phase singularity is defined as a point $\textbf{r}_i$ in the field $u(\textbf{r})$ such that $u(\textbf{r}_i)=0 \Longleftrightarrow \Re \{u(\textbf{r}_i)\}=\Im \{u(\textbf{r}_i)\}=0$ and a such, the field at $\textbf{r}_i$ has a undefined phase. To identify a polarisation singularity, we first begin by defining the Stokes parameters,
\begin{eqnarray}
    S_1(\textbf{r}) &=& \Re \{g(\textbf{r})u^*(\textbf{r})\}\,, \label{eq:StokesS1}\\
    S_2(\textbf{r}) &=& \Im \{g(\textbf{r})u^*(\textbf{r})\}\,,\label{eq:StokesS2}
\end{eqnarray}
where $S_1$ quantifies how horizontal/vertical the polarisation is and $S_2$ quantifies how diagonal/antidiagonal the polarisation is. With these parameters we can define the complex field $P=S_1 + iS_2$. A singularity in this field corresponds to a point $\textbf{r}_i$ where $P(\textbf{r}_i)=0 \Longleftrightarrow S_1(\textbf{r}_i)=S_2(\textbf{r}_i)=0$ and is characterised by a polarisation state of undefined ellipse orientation (i.e. the polarisation at that point is purely right circularly polarised or purely left circularly polarised).

The fundamental Gaussian mode is given by $g(\textbf{r}) = C e^{r^2/w_0^2}$  and so satisfies $|g(\textbf{r})|>0 \; \forall \; \textbf{r} \in \mathcal{R}^2$, where $C$ is a normalisation dependant constant. Combining this with Equations \ref{eq:StokesS1} and \ref{eq:StokesS2} implies that $S_1(\textbf{r}_1) = S_2(\textbf{r}_1) = 0$ if and only if $u(\textbf{r}_1) = 0$. Therefore, before entering an aberrated channel, the location of any polarisation singularities must be the same as the phase singularities embedded in the scalar field $u$. 

We next add a spatially varying phase perturbation $\Theta(\textbf{r})$ onto the initial vectorial field. We assume that the channel is one sided and so the perturbation affects both polarisations equally,
\begin{eqnarray}
    g &\rightarrow & g'  = ge^{i\Theta(\textbf{r})}\,, \\
    u &\rightarrow & u' = ue^{i\Theta(\textbf{r})} \,.
\end{eqnarray}
The phase aberration satisfies $|e^{i\Theta(\textbf{r})}|=1$ and thus does not add or remove any zeros from either of the two fields. As such, the phase singularities of $u'$ are located in the same positions as those of $u$ (i.e. $u'(\textbf{r}_i)=0 \; \forall \; \textbf{r}_i$ such that $u(\textbf{r}_i)=0$) and $g'$ does not have any zeros. Therefore, the only points $r_j$ that satisfy $S_1(\textbf{r}_j)=S_2(\textbf{r}_j)=0$ also satisfy $u'(\textbf{r}_j) = 0$. Therefore, the polarisation singularities must be located at the same positions as the phase singularities embedded in $u'$.

We begin to see phase singularity wander when the beams begin propagating. To examine the dynamics of the singularities in propagation, we begin with the Huygens-Fresnel diffraction integral.
\begin{equation}
    u'(\mathbf{r'},z) = \frac{e^{ikz}}{i\lambda z} \iint e^{ik(\mathbf{r}' - \mathbf{r})/2z} u(\mathbf{r},0) e^{i\Theta(\mathbf{r})} d^2\text{r}
\end{equation}
We can safely assume that we are working in the weak turbulence regime for two reasons. Fist, we simulate turbulence numerically and in the experiment with a single phase screen. Second, the far-field profiles of the aberrated Gaussian beams do not exhibit any emergent singularities. This means we can assume that $|i\Theta| \ll1$, allowing for the approximation $e^{i\Theta} \approx 1 + i\Theta$, where we ignore high order terms. Plugging this into the diffraction integral gives,
\begin{eqnarray}
    u'(\mathbf{r'},z) &=& u'(\mathbf{r'},z) + \frac{e^{ikz}}{\lambda z} \iint e^{ik(\mathbf{r}' - \mathbf{r}')/2z} u(\mathbf{r},0) \Theta(\mathbf{r}) d^2\text{r}  \,,\nonumber \\
    &=& u(\mathbf{r'},z) + D(\mathbf{r}',z)\,,
    \label{eq:FresenlPert}
\end{eqnarray}
where $u(\mathbf{r'},z)$ is the propagated unperturbed field in the measurement plane and $D(\mathbf{r}',z)$ represents an additive perturbation introduced by the distortions in the medium. Because we assume $|i\Theta| \ll1$, this immediately implies $u(\mathbf{r'},z) > D(\mathbf{r}',z)$ and thus the additive perturbation is not large enough to create any zeros in the propagated field. This means that the propagated Gaussian beam through the channel satisfies $|g'(\textbf{r}',z)| >0 \; \forall \;\textbf{r}' \in \mathcal{R}^2$ and so the position of a polarisation singularity $\textbf{r}_i$, where $S_1(\textbf{r}_i) = S_2(\textbf{r}_i) = 0$, can only exist for $\textbf{r}_i$ that also satisfy $u'(\textbf{r}_i,z) = 0$. Since these points are the positions of the phase singularities in $u'(\textbf{r},z)$, the phase and polarisation singularities must be co-located, even if the field has been propagated some distance.

We have therefore shown that the positions of the polarisation singularities is locked to the positions of the phase singularities at every point through the turbulent channel. This connection between the two singularities is due to the one sided nature of the channel. If the channel was two-sided and the polarisation DoF was also affected, perhaps through some birefringent object then the positions of both singularities would not be linked. This can be easily seen by applying some standard polarisation rotation $\hat{R}(\theta)$ (such as the action of a wave-plate) to both fields at the receiver. Before the wave-plate the singularities would be co-located, but in general after the wave-plate, the local polarisation state of the vectorial field at the singularity would have changed from purely right-circular $\hat{\textbf{e}}_\text{R} = (1,\; i)^T/\sqrt{2} $ to $\hat{R}(\theta)\hat{\textbf{e}}_\text{R}$. This point will no longer have an undefined ellipse orientation and thus would no longer be the location of the desired polarisation singularity. Because $\hat{R}(\theta)$ is unitary, the point of undefined ellipse orientation would have moved somewhere in the transverse plane, an extent determined by the magnitude of the rotation on the Poincar\'{e} sphere.

We have also made no assumptions about the exact structure of $u$ and thus the exact distribution of the phase singularities in the initial field. This means that for any possible configuration of polarisation singularities, there must exist some underlying distribution of phase singularities in one of the component scalar modes whose position is always linked to that of the polarisation singularities.

\section*{Supplementary: Full field measurement of scalar fields} 
To measure the complex transverse field of the scalar LG beam, we employ an interferometric acquisition followed by a spatial frequency selection \cite{gaasvik2003optical}.

The field under test is superposed with a coherent reference beam with a controllable phase ramp, The camera records a single intensity distribution containing interference fringes, whose spatial carrier shifts the field information away from the zero frequency component. This separation allows the isolation of one cross term that is linear in the unknown field, so that both amplitude and phase can be reconstructed without iterative procedures. 

In general, one can consider an unknown scalar at the camera plane to be
\begin{equation}
    U(x,y)=A(x,y)e^{i\varphi(x,y)},
\end{equation}
and the reference be
\begin{equation}
    R(x,y)=A_re^{i\varphi_r(x,y)},
\end{equation}
where $\varphi_r$ is chosen to include a transverse carrier. The measured interferogram is
\begin{equation}    
\begin{split}
    I(x,y)=&|U(x,y)+R(x,y)|^2\\
    =&|U(x,y)|^2+|R(x,y)|^2\\
    &+U(x,y)R^*(x,y) + U^*(x,y)R(x,y).
\end{split}
\end{equation}
Here, $^*$ denotes the complex conjugate. Taking the two dimensional Fourier transform of $I(x,y)$ yields a spectrum with a dominant low spatial frequency from $|U|^2+|R|^2$ and two displaced components associated with the cross terms $UR^*$ and $U^*R$. We select the displaced component that corresponds to $UR^*$ by applying a window aperture in the Fourier domain and then recentre it to the origin. The inverse Fourier transform of this recentred spectrum gives a complex quantity proportional to the unknown field $\tilde{U}(x,y)=CU(c,y)$, where $C$ is a proportionality constant. Now, the obtained complex field has amplitude and phase maps following respectively $A(x,y)=|\tilde{U}(x,y)|$ and $\varphi(x,y)=\text{arg}(\tilde{U}(x,y))$, up to a constant phase offset. In the context of this work, the perturbed field plays the role of the above unknown complex field. 

Selecting a single diffraction order requires cropping the Fourier spectrum around the chosen sideband, which reduces the sampled bandwidth and would also reduce the reconstructed spatial resolution after the inverse Fourier transform. To preserve the original camera resolution, the selected and recentred Fourier domain array is inserted into a larger array of zeros with the same size as the original Fourier transform prior to the inverse transform. This zero padding maintains the reconstruction on the native detector grid and yields a smooth spatial interpolation consistent with the camera resolution, while retaining the phase amplitude content carried by the isolated order. 

\section*{Supplementary: Stokes polarimetry and measuring polarisation phases}

\begin{figure}[htbp]
    \centering
    \includegraphics[width=0.45\textwidth]{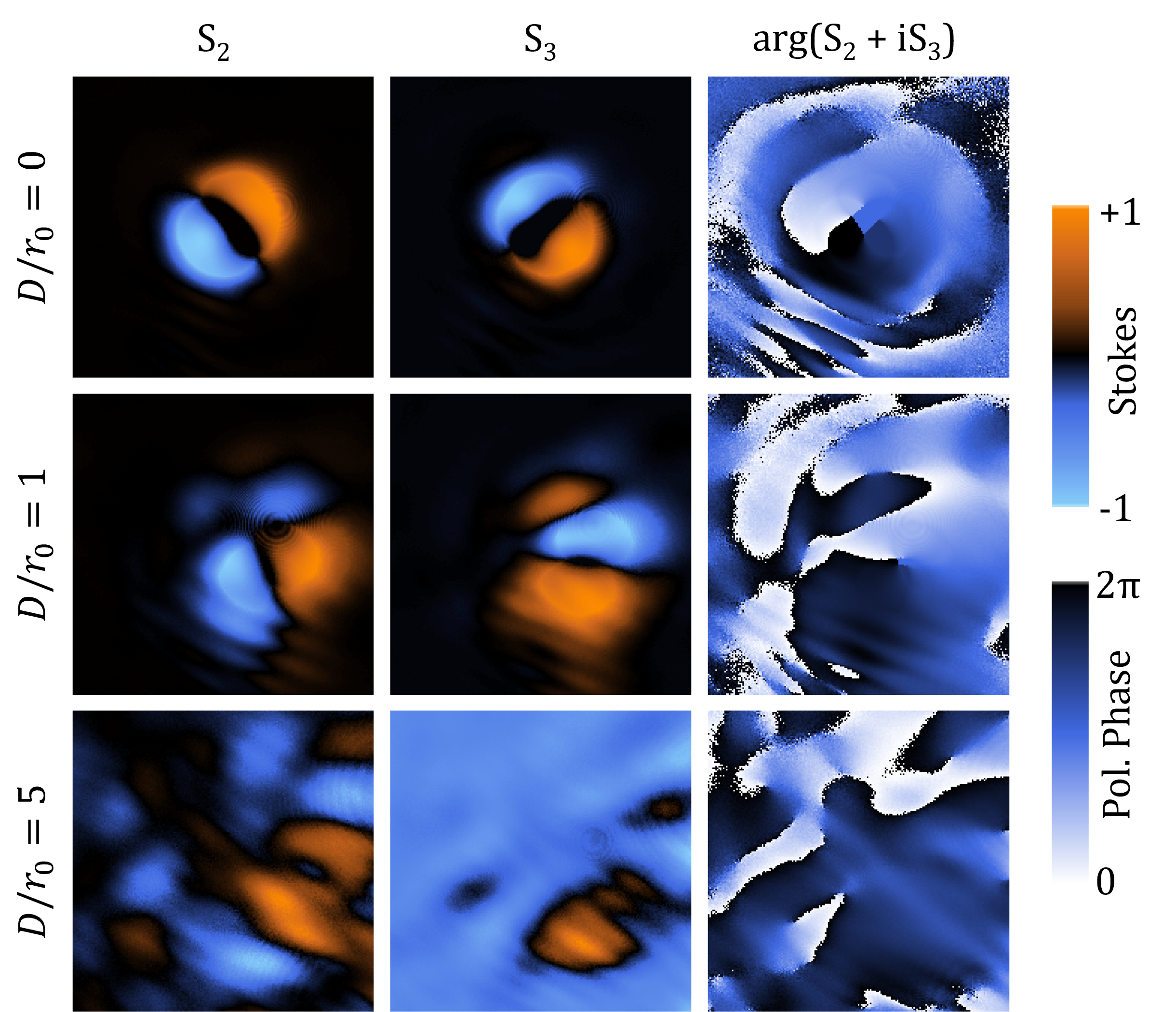}
    \caption{The experimentally measured Stokes parameters $S_2$ and $S_3$ and the polarisation phase calculated from them for various turbulence strengths.}
    \label{fig:Stokes}
\end{figure}

In this work, we generate vectorial, paraxial light fields whose polarisation and spatial degrees of freedom (DoFs) are non-separable creating spatially varying polarisation distributions\cite{rosales2018review}. The total field can be described as,
\begin{eqnarray}
        \psi_{\text{pol}}(\textbf{r}) &=&  E_x^*(\textbf{r})\hat{\textbf{e}}_\text{H}+ E_y(\textbf{r})\hat{\textbf{e}}_\text{V}, \nonumber \\
        &=& u_\text{LG}^{\ell_H}(\textbf{r}) \hat{\textbf{e}}_\text{H}+u_\text{LG}^{\ell_V} (\textbf{r}) \hat{\textbf{e}}_\text{V},        
        \label{eq:vector beam}
\end{eqnarray}
where $\hat{\textbf{e}}_{\text{H}(V)}$ represents the horizontal (vertical) polarisation directions, \textbf{r} is the transverse spatial coordinate and $u_\text{LG}^{\ell }$ are the LG spatial modes defined by Eq. \ref{eq:LGdef}. The polarisation state at each point in the field can be completely described using the Stokes vector, $\mathbf{S}(x,y)=[S_1,\; S_2,\; S_3]^T$ whose components are defined by \cite{born2013principles},
\begin{eqnarray}
S_1(\textbf{r} ) &=& E_x^*(\textbf{r} ) E_x(\textbf{r} ) -  E_y^*(\textbf{r} ) E_y(\textbf{r} ) \\ 
S_2(\textbf{r} ) &=& E_x^*(\textbf{r} ) E_y(\textbf{r} ) -  E_y^*(\textbf{r} ) E_x(\textbf{r} ) \\ 
S_3(\textbf{r} ) &=& i\left[E_y^*(\textbf{r} ) E_x(\textbf{r} ) -  E_x^*(\textbf{r} ) E_y(\textbf{r} )\right].  \label{eq:Stokes def}
\end{eqnarray}
The Stokes parameters can be easily obtained in the experiment through six polarisation intensity projections \cite{singh2020digital},
\begin{eqnarray} \label{eq:stokesS0}
    S_{0}(\textbf{r} ) &=& I_{H}(\textbf{r} ) + I_{V}(\textbf{r} ) \\
\label{eq:stokesS1} 
    S_{1}(\textbf{r} ) &=& I_{H}(\textbf{r} ) - I_{V}(\textbf{r} )\\
\label{eq:stokesS2}
    S_{2}(\textbf{r} ) &=& I_{D}(\textbf{r} ) - I_{A}(\textbf{r} )\\
\label{eq:stokesS3}
    S_{3}(\textbf{r} ) &=& I_{R}(\textbf{r} ) - I_{L}(\textbf{r} ) \,.
\end{eqnarray}
where the subscripts H, V, D, A, R and L represent horizontal, vertical, diagonal, antidiagonal, right circular and left circular polarisations respectively. The projections for $S_0$, $S_1$ and $S_2$ were obtained directly from the output of the polarisation sensitive camera and a quarter wave plate was used to obtain the circularly polarised intensity projections. Experimentally obtained Stokes parameters $S_2$ and $S_3$ are shown in the first two columns in Figure \ref{fig:Stokes} for $\ell_H = 0$ and $\ell_V = 1$ and for various turbulence strength.

Using the Stokes parameters, we may define complex fields whose amplitude and phase describe projections of the polarisation state at that point onto different planes through the Poincar\'{e} sphere, 
\begin{eqnarray} \label{eq:polphase1}
P_{1}(\textbf{r} ) &=& S_{2}(\textbf{r} ) + iS_{3}(\textbf{r} )\,,\\
\label{eq:polphase2}
P_{2}(\textbf{r} ) &=& S_{3}(\textbf{r} ) + iS_{1}(\textbf{r} )\,,\\
\label{eq:polphase3}
P_{3}(\textbf{r} ) &=& S_{1}(\textbf{r} ) + iS_{2}(\textbf{r} )\,.
\end{eqnarray}
For the purposes of defining and locating singularities, we will look at the argument of these complex fields, and restrict our study to the phase structure of $P_1$, due to its structural similarity in our chosen basis $\{\hat{\textbf{e}}_\text{H},\hat{\textbf{e}}_\text{V} \}$ to the phase of the electric field of OAM beams in unperturbed channels. Examples of the measured polarisation phases are shown in the third column of Figure \ref{fig:Stokes}.

\section*{Supplementary: Experimental setups and numerical simulations}

\begin{figure*}[htbp]
    \centering
    \includegraphics[width=0.9\textwidth]{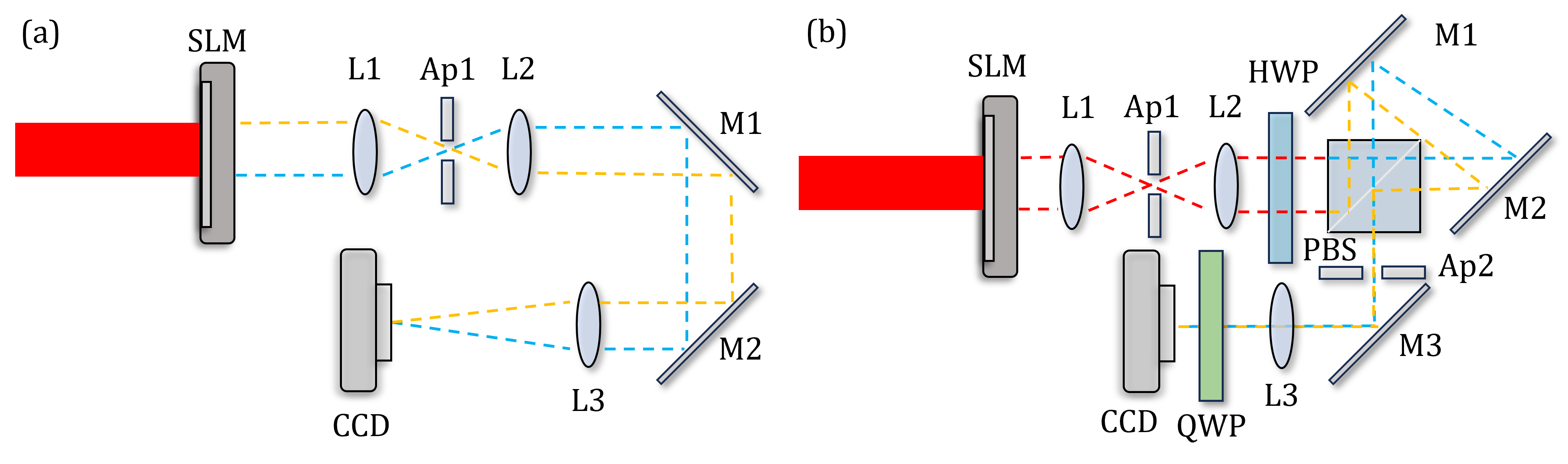}
    \caption{(a) The experimental set-up used to generate and measure the full-field of a scalar beam in the far-field after being aberrated by simulated atmospheric turbulence. (b) The experimental set-up used to generate vectorial beams and measure them in the far-field after being aberrated by simulated atmospheric turbulence. Here, SLM stands for spatial light modulator, L for lenses, Ap for iris apertures, M for mirrors, PBS for polarising beam splitter, HWP for half-wave plate, QWP for quarter-wave plate, CCD for the charge-coupled device camera.}
    \label{fig:Exp}
\end{figure*}

Experimentally, we generated optical phase singularities by constructing LG scalar light fields via the setup illustrated in Figure \ref{fig:Exp} (a). A horizontally polarised, collimated Gaussian beam from a HeNe laser (wavelength $\lambda = 633$~nm) was directed onto a Holoeye PLUTO 2.1 spatial light modulator (SLM). In addition to the LG mode modulation on the SLM, we included a reference Gaussian beam spatially displaced from the signal mode in the hologram to create an automated self-interfering arrangement. The hologram itself was computed using a complex amplitude modulation scheme \cite{arrizon2007pixelated}, 
\begin{equation}
    H(x,y) = J^{-1}_1 (A(x,y)) \sin[\phi(x,y) + 2\pi(G_x x + G_y y) ] \,,
\end{equation}
The signal mode and the reference were then imaged through a $4f$ filtering system composed of two lenses $L_1$ and $L_2$ ($f_1=f_2=300$~mm) and an aperture to isolate the first diffraction order in which the desired field is located. This plane is then imaged to the far field using a single lens $L_3$ of focal length $f_3 = 1000$~mm were a CCD (Allied Vision Mako G-508B POL) is placed to capture the final intensity. Due to the spatial displacement of the reference beam relative to the signal beam on the SLM, both beams interfere with a relative phase ramp in the far-field allowing for a full scalar field reconstruction using the method described earlier. As one should expect, the perturbed LG mode wanders across the transverse space at the far-field. To ensure consistent interference over many different realizations, we apply to the reference beam the first-order Zernike modes extracted from each turbulent phase screen. As these aberrations correspond to the vertical and horizontal tilts contributions \cite{noll1976zernike,niu2022zernike}, the interferometry is always satisfied for any phase screen applied to the signal beam. This technique is useful for experimental arrangements where interferometry is required while using a laser with short coherence length ($\mathtt{\sim}6$ cm for our HeNe laser).

The experimental setup for generating and detecting polarisation singularities is shown in Figure \ref{fig:Exp} (b). The same collimated Gaussian beam as before was incident onto the same SLM. In this instance, the SLM screen was divided into two regions, one encoding a Gaussian mode, and the other a Laguerre-Gaussian (LG) mode carrying nonzero topological charge $\ell$. After passing through the same 4f filtering system, the beams pass through a half wave plate (HWP), converting them to diagonal polarisation. They subsequently entered a modified Sagnac interferometer consisting of a polarising beam splitter (PBS) and two mirrors, creating a coaxial and co-propagating vectorial superposition according to Equation \ref{eq:vector beam}. The vector beam is then imaged to far-field using a the same Fourier transforming lens as before. The polarisation sensitive camera and quarter wave plate (QWP) were then used to perform an over-complete set Stokes polarimetry measurements to fully characterise the polarisation profile of the beam. Identical phase screens are endocded onto each half of the SLM to ensure both vectorial components experience the exact same distortion.

Our numerical algorithm is designed to closely mirror the experimental sampling and propagation, so the same analysis can be applied without additional rescaling assumptions. First, the scalar and vector beams are synthesized on two-dimensional arrays matching the pixel resolution from the SLM ($1920\times1080$ and 8 $\mu$m pixel pitch), ensuring that the discretized beam shaping in simulation reproduces the experimentally imposed sampling. Then, a thin-lens transfer function corresponding to a 1000 mm focal length is applied to the generated field, finally propagated to the focal position using the angular spectrum method \cite{goodman2005introduction}. After propagation, the simulated field is re-mapped onto a second transverse grid via interpolation, chosen to match the effective sampling of the polarisation-resolving camera used in the measurements ($2464\times2056$ and pixel pitch of 3.45 $\mu$m). This is essential to preserve the camera-limited spatial resolution when comparing the singularity spatial statistics with experiments. With the propagated field distributions expressed on a consisted grid, we apply the same centre-of-mass procedure and singularity-tracking approach previously described to build the singularity maps and extract the corresponding wandering total deviation radii. 

\section*{Supplementary: Singularity tracking in complex fields}

\noindent With the phase structure of both the scalar and vectorial fields measured, it now remains to locate and track their respective singularities. In ideal conditions, one may simply search for positions where the real and imaginary components of the field of interest equal zero. However, the presence of noise in experimental data makes this approach prone to error. We therefore implement a numerical procedure first proposed by Chen \textit{et. al.} \cite{chen2007detection}, which leveraged a numerical implementation of the curl operation.

We begin with an arbitrary phase $\Theta(x,y)$, that can be written as a sum of a continuous phase function $\theta(x,y)$ and the phase profiles of an arbitrary number $n$ of canonical singularities $\phi(x,y)$,
\begin{equation}
    \Theta(x,y) = \theta(x,y) + \sum_n \phi(x-x_n,y-y_n)\,.
    \label{eq:GeneralPhase}
\end{equation}
Analytically the singularity charge $\tau$ is determined by computing the contour integral over the gradient of the phase function along a closed path $\mathcal{C}$ that encircles the singularity of interest,
\begin{equation}
    2\pi\tau = \oint_\mathcal{C} \nabla\Theta(x,y) \cdot \text{d}l\,,
\end{equation}
Experimental data cannot be known analytically and so we turn an alternative approach. We define the gradient of the measured phase as $\mathbf{G}(x,y) = \nabla\Theta(x,y)$ where $\mathbf{G}(x,y) = G_x(x,y)\hat{x} + G_y(x,y)\hat{y}$ is a vector field. The gradient $\mathbf{G}(x,y)$ has no $z$-dependence and has no $z$-component, meaning the curl of this function is a scalar value, representing the $z$-component of the curl,
\begin{equation}
    \nabla \times \mathbf{G}(x,y) = \left( \frac{\partial G_x(x,y)}{\partial y} - \frac{\partial G_y(x,y)}{\partial x} \right) \hat{z}\,.
    \label{eq:GradCurl}
\end{equation}
For the continuous component, the curl of the gradient will vanish (i.e., $\nabla\times \nabla\theta = 0 $), but will be non-zero for the singularity component,
\begin{equation}
    \nabla \times \nabla \phi(x,y) = 2\pi\tau\delta(x)\delta(y).
\end{equation}
The curl of the gradient of Equation~\ref{eq:GeneralPhase} then becomes,
\begin{equation}
    \nabla\times\nabla\Theta(x,y) = 2\pi \sum_n\tau_n\delta(x-x_n)\delta(y-y_n)\,.
\end{equation}
We can therefore immediately determine the charge of any singularity $\tau_n$ and its location $(x_n,y_n)$. Simulated an experimental data is all defined in discrete arrays making a direct analytical computation of the curl prone to errors. Instead, we make use of a numerical approximation of Equation~\ref{eq:GradCurl} known as the circulation $D$,
\begin{equation}
    D^{m,n} = G_x^{m,n} - G_x^{m+1,n} +G_y^{m,n} -  G_y^{m,n+1}\,.
    \label{eq:circulation}
\end{equation}    
Here, $D^{m,n}$ represents the value of the circulation of the pixel in the $n$-th row and $m$-th column. $G_x^{m,n}$ and $G_y^{m,n}$ are the phase gradients in the horizontal and vertical direction of the pixel in the $n$-th row and $m$-th column, respectively. Typically, the circulation will return a non-zero value if there is a singularity at a particular pixel and a non-zero value otherwise. The magnitude of the circulation will be approximately $2\pi\tau$, where $\tau $ indicates the charge of the singularity, and the sign indicates the direction/handedness of the singularity.

\section*{Supplementary: Simulating turbulence with digital holograms}

\begin{figure*}[htbp]
    \centering
    \includegraphics[width=0.9\textwidth]{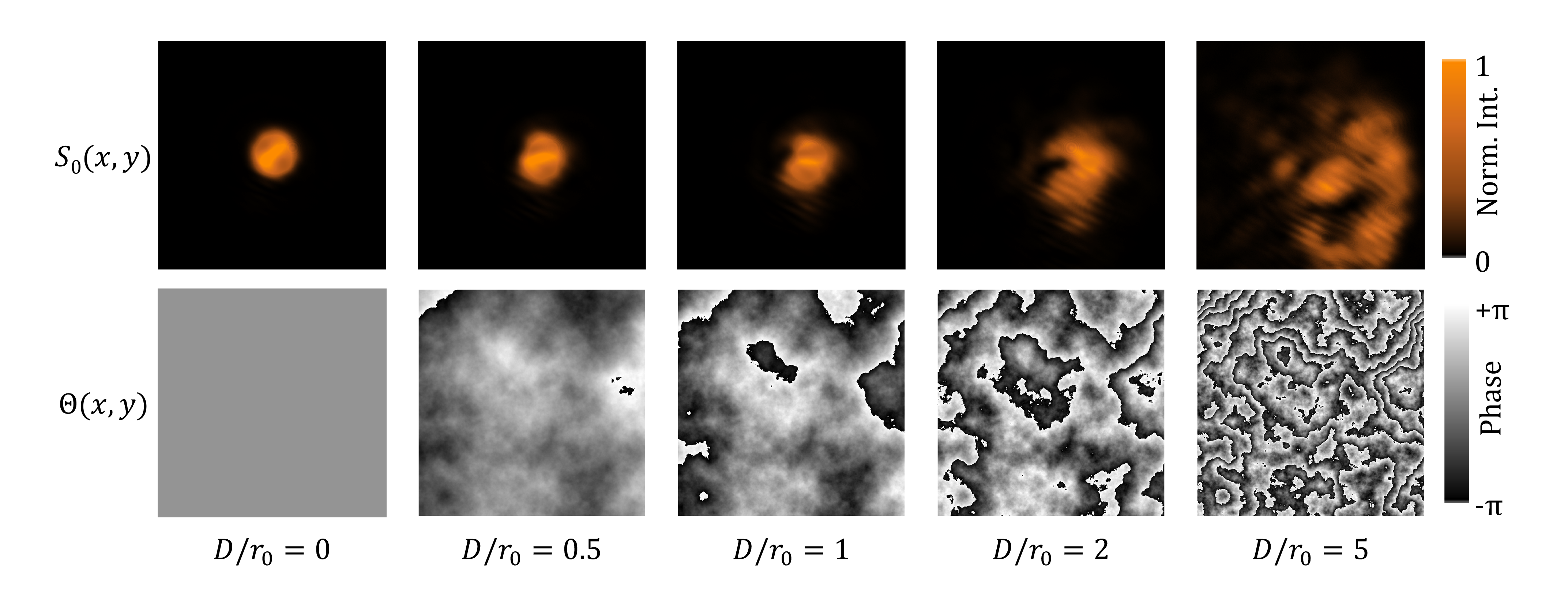}
    \caption{The experimentally measured intensities and the corresponding turbulence phase screen used to perturb the incident complex fields.}
    \label{fig:TurbInt}
\end{figure*}

In order to emulate the effects of atmospheric turbulence in a controlled manner, we leverage the power and versatility of digital holograms to encode phase screens onto our beams. The phase screens are generated to replicate the refractive index distortions that a complex light field would experience when travelling through a real-world turbulent channel. The full theory, generation details and code for generating the screens can be found in Ref. \cite{peters2025structured}.

We opted to use the Fourier phase screen generation method due to its computational efficiency. The method randomly samples frequency coefficients from a normal distribution of mean zero and unit variance. The coefficients are then scaled according to a model specific power spectral density (PSD) function. The degree of this scaling depends on the desired distortion strength. In order to accurately capture the low frequency contributions of turbulence, we implement the technique of subharmonic sampling up to a maximum of 5 subharmonics.

While there are many possible PSD functions that could be used, we chose the Kolmogorov PSD, $\Phi(k)$, as its analytical behaviour has been studied in many contexts across the literature. Its is given by,
\begin{equation}
   \Phi(k) = 0.023 r_0^{-5/3}k^{-11/3}\,,
   \label{eq:kol PSD}
\end{equation}
where $k$ represent the magnitude of the angular frequency coordinate $k=\sqrt{k_x^2+k_y^2}$ and $r_0$ is the Fried parameter which represents the average correlation length of phase distortion. Examples of turbulence phase screens are shown in Figure \ref{fig:TurbInt}, along with the corresponding far-field intensity profiles. Weaker phase screens are characterized by a large Fried parameter  while stronger phase screens are characterized by smaller Fried parameter.

The strength of the distortion experienced by a beam depends on not only on the $r_0$ of a particularity channel, but also on the transverse spatial extent of that beam relative to $r_0$. We therefore quantify the turbulence strength using unitless parameter $D/r_0$, where $D$ is the beam diameter.This means we can easily categorise the strength of a given channel in weak when the Fired parameters is much larger than the beam size ($D/r_0<1$), moderate when the Fried parameter is approximately equal to the beam size ($D/r_0\approx1$) and strong when it is smaller than the beam size (($D/r_0>1$)). In the case of unnormalised beam sizes, we set $D=2w_0$ and for normalised beam size we set $D=2w$.

\section*{Supplementary: Extracting first-order Zernike polynomials from turbulence phase screens}

\begin{figure}[htbp]
    \centering
    \includegraphics[width=0.4\textwidth]{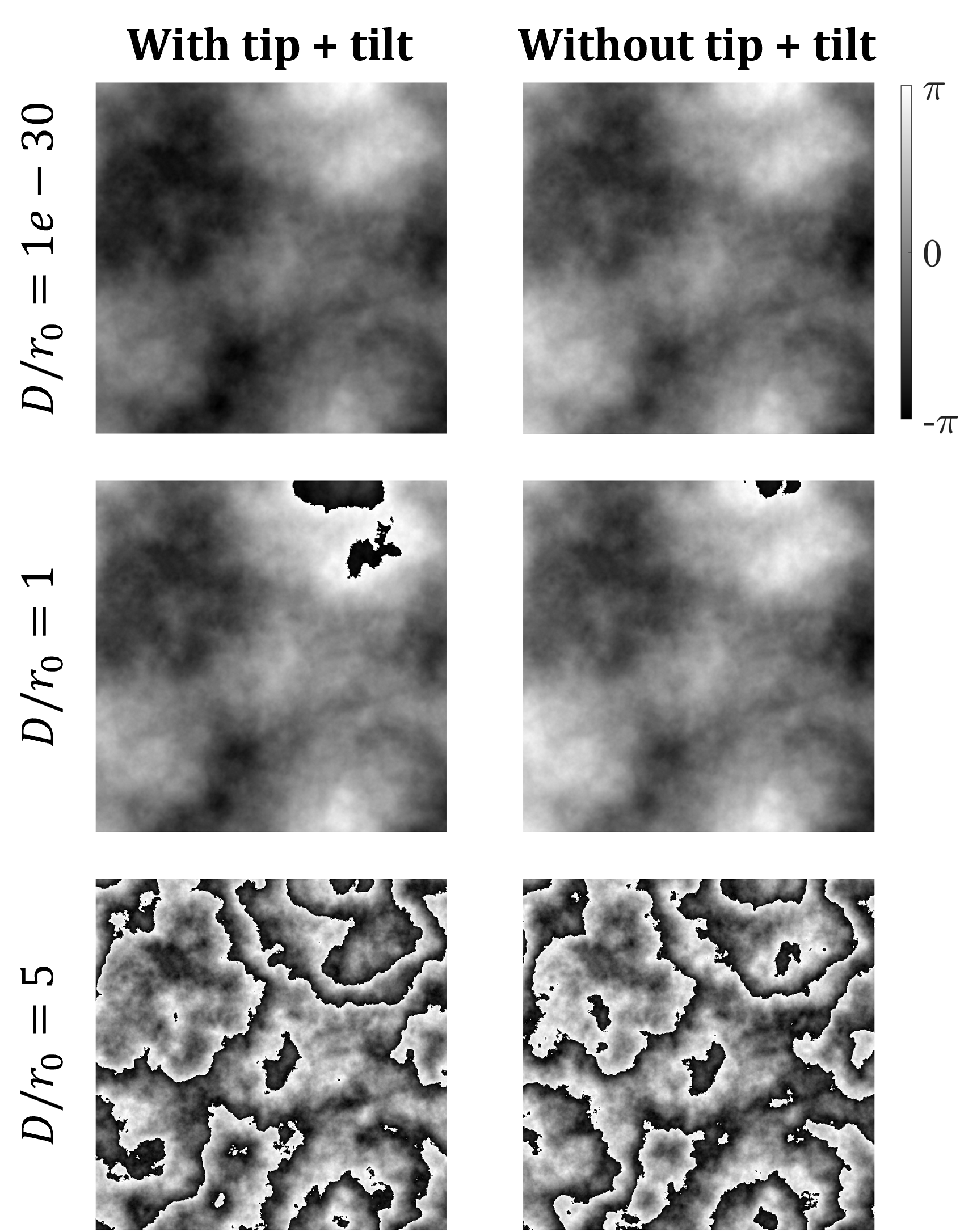}
    \caption{Examples of turbulence phase screens of varying turbulence strengths before and after removing the tip and tilt contributions.}
    \label{fig:TipTilt}
\end{figure}

Atmospheric turbulence can also be well described in terms of Zernike polynomials \cite{noll1976zernike}, which reveals that the polynomials corresponding to the tip and tilt aberrations (vertical and horizontal phase ramps) constitute a majority of the experienced distortion. Since these aberrations will affect the final position of both the phase and polarisation singularities in the exact same way, we remove them from their contritions from our generated phase screens. This is easily permitted due to the orthogonality of the Zernike polynomials. This allows us to isolate and study only the effects of high order aberrations on the stability of the singularities. This corrected phase screen $\Theta_{\text{corrected}}(x,y)$ is computed as follows,
\begin{equation}
    \Theta_{\text{corrected}} = \Theta - \langle Z_1^1|\Theta \rangle - \langle Z_1^{-1}|\Theta \rangle\,,
\end{equation}
where $\Theta$ is the uncorrected phase screen, $Z^1_1$ and $Z^1_{-1}$ are the Zernike polynomials corresponding to tip and tilt respectively and $\langle \cdot|\cdot \rangle$ represents the inner product. Examples of phase screen before and after tip and tilt correction are shown in Figure \ref{fig:TipTilt}.

\section*{Supplementary: Calibration of turbulence phase screens}

\begin{figure}[htbp]
    \centering
    \includegraphics[width=0.4\textwidth]{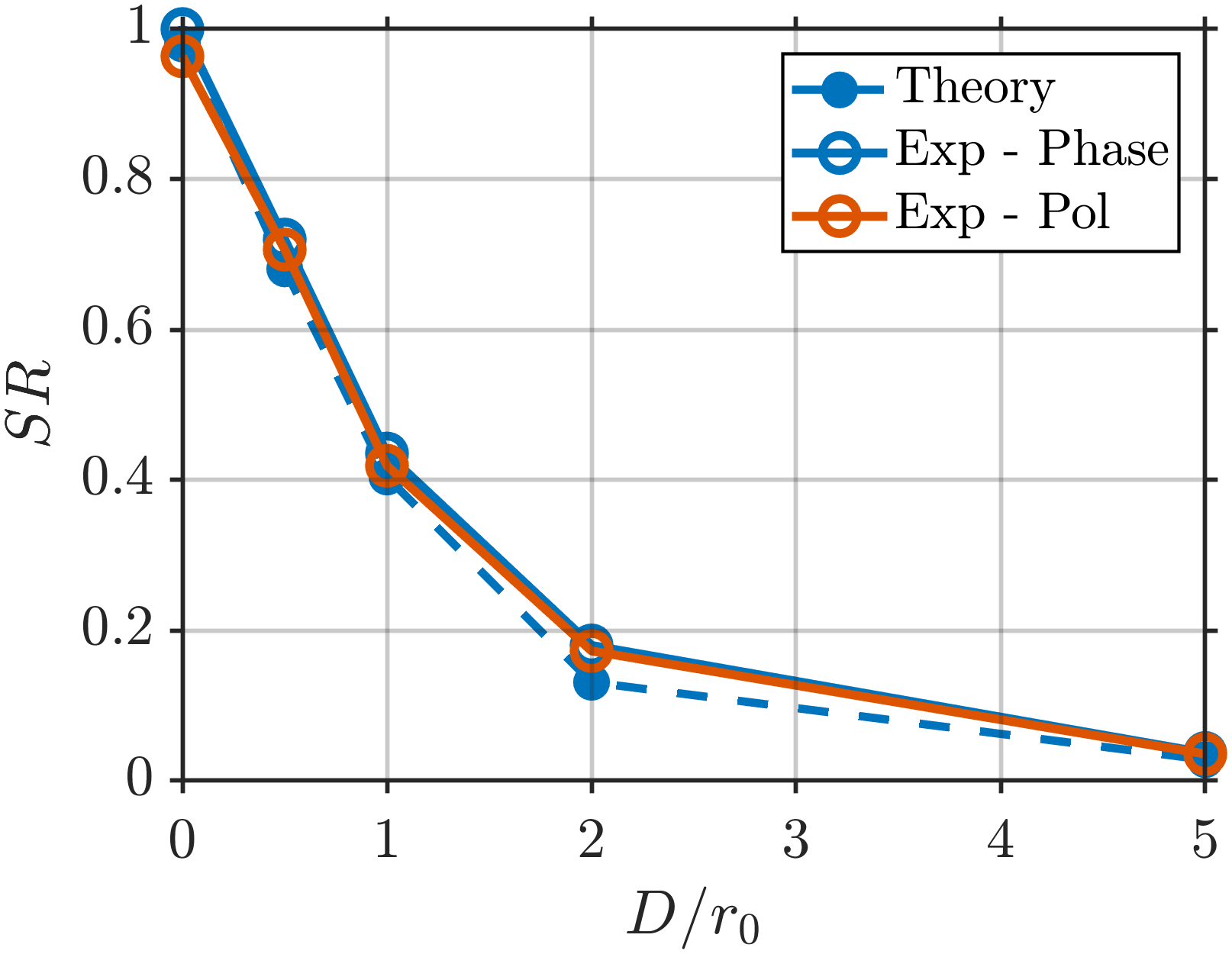}
    \caption{Experimentally measured Strehl ratios as a function of turbulence strength for the phase singularity experimental setup and the polarisation singularity experimental measurement.}
    \label{fig:StrehlCal}
\end{figure}

The experimental apparatus used to generate and detect phase singularities differed from that used to generate and detect polarisation singularities. As such, calibration tests were performed on both set-ups to ensure that the same distortions strengths were encoded in both instances. To perform this calibration, we make use of the a metric typically referred to as the Strehl ratio,
\begin{equation}
    SR = \frac{\langle I(0,0,t) \rangle}{I_0(0,0)}
\end{equation}
quantifying the time (or phase screen) averaged on axis intensity $\langle I(0,0,t) \rangle$ of a distorted incident beam in relation to the on axis intensity in the absence of turbulence $I_0(0,0)$. For the Kolmogorov PSD, the $SR$ has an explicit dependence on the turbulence strength given by,
\begin{equation}
    SR \approx \frac{1}{ \left[1 + (D/r_0)^{5/3} \right]^{6/5}} \, .
\end{equation}
Figure \ref{fig:StrehlCal} shows the measured Strehl ratio as function of turbulence strength $D/r_0$ for both the phase singularity apparatus and the polarisation singularity apparatus, and we observe almost perfect agreement to theory in both cases. Each data point represents the average over 100 independent phase screen realizations.

\end{document}